# Inorganic photonic materials for lasers and biomedicine in the infrared

*H.-T. Sun*

*Hokkaido University, Sapporo 060-8628, Japan*

Inorganic photonic materials demonstrating luminescence in the infrared spectral range (> 1 μm) play a vital important role in modern information, telecommunication, and biomedicine disciplines. For example, infrared luminescent nanoparticles occupy an important position in scientific research because of their wide range of potential applications. In optoelectronics area, with the rapid development of telecommunication technology, the conventional $Er^{3+}$-doped silica fiber amplifier has difficulty in meeting information transportation requirements. An attractive way to meet these requirements is to expand the number of transmission channels, which strongly depend on the gain bandwidth of the amplifier and the laser source. Therefore, the development of broadband NIR amplifiers and optical sources to cover the 1.2~1.6μm telecommunication windows becomes a key objective to permit the whole optical windows of silica to be utilized and to achieve an efficient wavelength-division multiplexing transmission network. In recent years, solution-processed colloidal IV-VI (*PbS and PbSe*) nanocrystals or quantum dots (QDs) embedded in a semiconducting polymer matrix present a promising route to realize above purpose.[1] Although optical amplification and lasers, with emission energies tunable in the NIR, have been demonstrated for PbS and PbSe nanocrystals, however, the intrinsic drawbacks for this type of material makes them nearly impossible for large scale applications. This is based on the following facts: (i) Silica is the dominant material used in the modern telecommunications. Obviously, the use of PbS and PbSe based materials would largely increase the insertion loss for the optical devices; (ii) The toxicity of these lead-containing materials would greatly, if not completely, dwarf their commercial future owing to the strict environmental regulations worldwide. Concurrently, in biomedicine area, NIR optical imaging of living tissue by using QDs is an area of growing interest. Deep-tissue imaging requires the use of NIR light within a spectral window separated from the major absorption peaks of hemoglobin and water. Theoretical modeling studies have indicated that two spectral windows are available for *in vivo* imaging (2nd one: 1025-1150 nm & 3rd one: 1200-1400 nm).[2] Most recently, NIR emitting CdTe (CdSe) core (shell) QDs have been used as stable fluorescent tracers for lymph node mapping in living animals. In comparison with organic dyes, the main advantage of QDs resides in their resistance to bleaching over long periods of time (minutes to hours), allowing the acquisition of images that are well contrasted. However, the extreme toxicity of these cadmium-containing products makes them nearly impossible for the practical applications in humans,[3] although some animal experiments were carried out. In fact, the European Union has already issued a policy banning all of the potential products with any amount of intentionally added cadmium. As is well known, silica is biocompatible and ''generally recognized as safe'' by the US Food and Drug Administration (FDA). Thus, nanostructured silica materials have been investigated for the application as drug delivery vehicles in biomedicine region. If optically functionalized silica nanoparticles can emit in the 2nd and 3rd biological optical windows (1000-1400 nm), *it is hopeful to use this kind of toxic-free nanostructures as the reporters of NIR in vivo imaging.*[4]

Generally, the inorganic material systems used for infrared photonics can be classified into three groups, i.e., glasses, crystals, and hybrid structures such as superlattices. Among them, glasses represent one of the most important materials, which have already found a broad range of applications in commercial areas. Snitzer's first glass laser in 1961 spawned many new inventions in the field of optical devices and fiber communications.[5] In particular, optical glasses and devices performing in the infrared spectral ranges over 1 μm have attracted extensive attention over the past decades, because of their broad practical applications for telecommunications, spectroscopy, chemical and bimolecular sensing, imaging, integrated optics, materials processing and homeland defense. Generally, the materials and devices can be classified into passive and active ones, and many new techniques have been developed for their fabrication, characterization and functional applications. Up till now, diverse glass materials including silicate,[6] phosphate,[6] fluorophosphates,[7,8,9,10] germanate,[11] tellurite,[12,13,14] bismuthate,[15,16,17] fluoride,[18] oxyfluoride,[19] and chalcogenide[20] have been intensively studied to meet such requirements. It is necessary to note that glasses containing heavy metal oxide and halide, especially lead oxide and halide, can seriously destroyed the crucibles used, thus strongly influencing the designed material compositions. In addition, melting glasses containing elements with multitype valence also has a strict requirement for the melting atmosphere and crucibles.[15,21] Another notable barrier for the application of non-silica based glasses lies in their compatibility with mature technology widely employed. Thus, a number of topics occur in this exciting area, which is summarized as follows.

(1) New synthesis and fabrication approaches of infrared photonic glasses for high-strength optical windows, lasers, lenses and nonlinear optics in infrared spectral regions;

(2) Rare earth doped glasses, films and waveguides for near- and mid-infrared amplifiers and fiber lasers;

(3) Rare earth doped silica/silicon based glassy films and waveguides, and their applications for amplifiers and energy harvesting and conversion;

(4) New fabrication techniques of passive silicon-compatible glassy films and waveguides for integrated photonics;

(5) Experimental and theoretical design, fabrication, structural and optical characterization of new kinds of active photonic glasses, films and fibers.

Optical crystals are important for laser science. The most successful application of optical crystals is the Ti:sapphire, which has become a general and widely-adopted laser source in many optics and laser material labs. To realize stable performance of active crystals in devices, it is of paramount importance in the control of crystal qualities. Concurrently, hybrid materials such as superlattices consisting of alternative glass and crystalline layers have been investigated by many groups,[22,23] with aims to manipulate active centers and the overall performance of the devices built. In this regard, there are some interesting topics to be explored.

(1) The fabrication techniques for large-size bulk crystals to realize low-cost application;

(2) Exploring new optical crystals activated by some peculiar active centers such as bismuth;

(3) New wavelength laser sources based on the revolution of optically active crystals;

(4) Low-cost miniature lasers and high-power lasers.

Future infrared photonics greatly rely on the development of inorganic photonic materials, not only via simple composition optimization, but also based on the great improvement of the design, fabrication techniques, and functional integration. With enough efforts, it is believed that novel inorganic materials could offer abundant attractive properties, which make them an excellent choice for infrared science and nonlinear optics.

5